\documentclass[twocolumn]{aastex6}
\usepackage{epsfig}
\usepackage{graphicx}
\usepackage{float}
\usepackage{amsmath}
\usepackage{color}
\usepackage{amssymb}
\usepackage{amsfonts}
\usepackage{units}
\usepackage{bm}

\def\be{\begin{eqnarray}}
\def\ee{\end{eqnarray}}

\begin{document}

\title{Polarization Predictions in the GRB Prompt Phase with the Internal Shock Model}

\author{Mi-Xiang Lan$^{1}$, Hao-Bing Wang$^{1}$, Siyao Xu$^{2}$, Siming Liu$^{3}$, and Xue-Feng Wu$^{4}$}
\affil{$^{1}$Center for Theoretical Physics and College of Physics, Jilin University, Changchun, 130012, China; lanmixiang@jlu.edu.cn \\
$^{2}$Institute for Advanced Study, 1 Einstein Drive, Princeton, NJ 08540, USA \\
$^{3}$Key Laboratory of Dark Matter and Space Astronomy, Purple Mountain Observatory, Chinese Academy of Sciences, Nanjing 210023, China \\
$^{4}$Purple Mountain Observatory, Chinese Academy of Sciences, Nanjing 210023, China; xfwu@pmo.ac.cn\\}

\begin{abstract}
As the standard gamma-ray burst (GRB) prompt-emission model, the internal shock (IS) model can reproduce the fast-rise and slow-decay features of the pulses in the GRB light curve. The time- and energy-dependent polarization can deliver important physical information on the emission region and can be used to test models. Polarization predictions for the GRB prompt phase with the magnetized IS model should be investigated carefully. The magnetic field of the magnetized IS model is very likely to be mixed and decays with radius. The synchrotron emission in the presence of such a decaying magnetic field can recover the Band-like spectrum of the GRB prompt phase. We investigate the dependence of the polarization of GRB prompt emission on both time and energy in the framework of the magnetized IS model. Due to the large range of parameters, it is hard to distinguish the magnetized IS model and the magnetic-reconnection model through polarization degree (PD) curves. The energy-dependent PD could increase toward the high-energy band for the magnetized IS model, while it decreases to zero above the megaelectronvolt band for the dissipative photosphere model. Therefore, we conclude that the energy dependence of PD can be used to distinguish these two models for the GRB prompt emission. Finally, we find that, independent of the observational energy band, the profiles of the $\xi_B-PD$ curve for the time-integrated and time-resolved PDs are very similar, where $\xi_B$ is the magnetic field strength ratio of the ordered component to the random component.
\end{abstract}


\section{Introduction}
Gamma-ray bursts (GRBs) are the most intensive explosions in the universe. Several thousands of observations have been accumulated, about two dozen of which have been reported with polarization detections (Yonetoku et al. 2011,2012; Covino \& G\"{o}tz 2016; Chattopadhyay et al. 2019; Zhang et al.2019; Kole et al. 2020). The observed polarization degrees (PDs) of GRB prompt emissions show rich diversity, ranging from a few percent to several dozens (Yonetoku et al. 2011,2012; Covino \& G\"{o}tz 2016; Chattopadhyay et al. 2019; Zhang et al.2019; Kole et al. 2020). The time-integrated PDs concentrate around $10\%$ for POLAR's detections (Zhang et al. 2019; Kole et al. 2020), while they are larger than $50\%$ for most GRBs detected by GAP (Yonetoku et al. 2012) and AstroSat (Chattopadhyay et al.2019). The time-resolved polarization angles (PAs) show signatures of evolution within pulses (Burgess et al. 2019; Zhang et al. 2019) and between pulses (Yonetoku et al. 2011). Although great effort has been made in the gamma-ray polarization detection, the present data still have large errors, which make it very difficult to distinguish models of the GRB prompt phase.

Some GRB pulses show fast-rise and slow-decay features (Bhat et al. 1994), which can be reproduced by the internal shock (IS) model (Kobayashi et al. 1997). In the scenario of IS (Pacz$\acute{y}$nski \& Xu 1994; Rees \& M$\acute{e}$sz$\acute{a}$ros 1994), the outflows launched from the GRB central engine will be highly variable. The faster shell will catch up and collide with the slower shell. Then shocks are formed, and the shock-accelerated electrons will radiate the GRB prompt emission. Literally, most of the GRB's spectrum can be described by a Band function (Band et al. 1993), which indicates a nonthermal origin of the photons. The low-energy and high-energy spectral indexes of the $\nu f_{\nu}$ spectrum are usually 1 and 0, respectively (Preece et al. 2000). The break energy of the Band spectrum is typically at 250 keV. It was pointed out that the synchrotron emission in a decaying magnetic field can reproduce the Band-like spectrum of the GRB prompt phase (Uhm \& Zhang 2014).

Waxman (2003) proposed a geometric model with the observational angle $\theta_V$ slightly outside the jet cone, ranging from $\theta_0$ to $\theta_0+1/\Gamma$, where $\theta_0$ and $\Gamma$ are the half-opening angle and the bulk Lorentz factor of the jet, respectively. Lyutikov et al. (2003) found up to $50\%$ PD can be achieved for the electromagnetic model (Lyutikov \& Blandford 2002,2003), which gives a natural origin of the large coherent magnetic field in the jet. It was pointed out by Granot (2003) that in producing a large PD, a transverse ordered magnetic field in the shock plane is the most likely case compared with a magnetic field normal to the shock plane and with a random magnetic field confined in the shock plane.

Time-integrated and energy-integrated polarization will erase the evolution information of polarization, which will be important for constraining the physical process of the GRB prompt phase. For example, a gradually evolving PA with time or energy will indicate an aligned magnetic field component in the jet and correspond to a magnetar-driven GRB (Lan et al. 2016; 2019). Recently, time-resolved polarization of GRB 170114A was analyzed by Burgess et al. (2019). Although the errors are very large, an increasing PD and a rotating PA were found during the burst. Then, time-resolved and energy-dependent polarization was investigated in the framework of a magnetic-reconnection model (Lan \& Dai 2020). Cheng et al. (2020) discussed both the time-resolved and time-integrated polarizations of the GRB prompt phase with a decaying magnetic field. The IS model, as one of the most popular models of the GRB prompt phase, and its time-evolved and energy-dependent polarization should be studied.

In this paper, we have investigated the polarization properties of GRB prompt emission with both time and spectrum in the framework of the magnetized IS model. The paper is arranged as follows. In Section 2, we present our model and numerical results. Finally, we give our conclusions and discussion in Section 3.

\section{The Model and The Numerical Results}
\subsection{Dynamics}
The basic picture of the IS model is that different shells with various velocities collide with each other and then the shocks are formed. Here, the magnetized IS (Fan et al. 2004) is considered. We denote the masses of the shells as $m_r$ and $m_s$, Lorentz factors as $\gamma_r$ and $\gamma_s$, the widths as $l_r$ and $l_s$ and the magnetization parameters as $\sigma_r$ and $\sigma_s$. The subscripts $"r"$ and $"s"$ represent the rapid shell and the slower shell, respectively. The initial separation of the two shells is set to be $l_{sep}$. The fast shell will catch up and collide with the slower shell ahead of it. Two shocks are formed, with the reverse shock (RS) propagating into the rapid shell and the forward shock (FS) into the slower one. Four regions are separated by two shocks: (1) the unshocked slower shell, (2) the shocked slower shell, (3) the shocked rapid shell, and (4) the unshocked rapid shell. The dynamics of the magnetized IS can be found by solving four nonlinear equations (Equations (19)-(22) in Fan et al. (2004)). Magnetic dissipation is not considered here.

The RS (FS) crossing time is $t_{rs}=l_r/c(\beta_r-\beta_{rs})$ ($t_{fs}=l_s/c(\beta_{fs}-\beta_{s})$), where $c$ is the speed of the light, and $\beta_r$, $\beta_s$, $\beta_{rs}$ and $\beta_{fs}$are the dimensionless velocity of the rapid shell, the slower shell, the RS, and the FS, respectively. The crossing radii of the RS and FS are $R_{rs}=\beta_{rs}ct_{rs}+R_{col}$ and $R_{fs}=\beta_{fs}ct_{fs}+R_{col}$. The collision radius is expressed as $R_{col}=\beta_rct_{col}$, with the collision time $t_{col}=l_{sep}/c(\beta_r-\beta_s)$.

In the GRB prompt phase, the shallowly decaying part of the light curve is often thought to be due to the high-latitude emission of the jet. Therefore, the integral of the jet emission should be made on the equal arrival time surface (EATS), and it can be expressed as follows:
\begin{equation}
t-t_{col}=\frac{t_{obs}/(1+z)-(1-\cos\theta)R_{col}/c}{1-\beta_f\cos\theta}
\end{equation}
The first photon that is detected by the observer comes from the shell at $R_{col}$. Here, $t$ is the burst source time, $t_{obs}$ is the observational time, $z$ is the redshift of the source, $\beta_f$ is the dimensionless velocity of the shocked regions, and $\theta$ is the angle between the velocity of the jet element and the line of sight. No lateral expansion is assumed. Since there is a minimum radius of the emission region ($R_{col}$), it will be a maximum value $\theta_{max}$ in one EATS and $\cos\theta_{max}=1-ct_{obs}/[(1+z)R_{col}]$.

\subsection{Shock Microphysics}
The number density of region 4 is $n'_4=m_r/[2\pi(1-\cos\theta_0)R^2l_r\gamma_rm_p]$. Here, $R=R_{col}+\beta_fc(t-t_{col})$ is the radius of the emission region, and $m_p$ is the mass of the proton. In this paper, the primed quantities are expressed in the comoving frame. For the magnetized shock, the number density of region 3 is $n'_3=u_{4s}n'_4/u_{3s}$, where $u_{js}=\gamma_{js}\beta_{js}$ ($j=3,4$), $\gamma_{js}$ is the Lorentz factor of region j measured in the shock frame, and the corresponding dimensionless velocity is $\beta_{js}$. The internal energy density of region 3 can be expressed as follows (Fan et al. 2004; Zhang \& Kobayashi 2005):
\begin{equation}
e'_3=n'_3m_pc^2(\gamma_{34}-1)\left(1-\frac{\gamma_{34}+1}{2u_{3s}u_{4s}}\sigma_r\right)
\end{equation}
where $\gamma_{34}$ is the Lorentz factor of region 3 measured in region 4. The electrons are accelerated in the shock and are injected as a power-law distribution with an index of $p_3$. The minimum Lorentz factor of shock-accelerated electrons is
\begin{equation}
\gamma_{m,3}=\frac{p_3-2}{p_3-1}\epsilon_{e,3}\frac{e'_3}{f_3n'_3m_ec^2}+1
\end{equation}
where $\epsilon_{e,3}$ is the energy equal participation factor of electrons in region 3, and $m_e$ denotes the mass of an electron. To match with the observed peak energy $E_{p}$ of the $\nu f_{\nu}$ spectrum, part $f_{3}$ of the electrons in the shocked regions are accelerated to the relativistic velocities (Daigne et al. 2009).

Turbulence plays an important role in the magnetic-reconnection model of GRB prompt emission (Zhang \& Yan 2011; Lazarian et al. 2019). As is shown in Deng et al. (2017), the ordered magnetic field with magnetic parameter $\sigma<10^{-3}$ will be destroyed by density perturbation, while its configuration will be retained with a moderately magnetic parameter (e.g. $\sigma=0.1$). In our following calculation, we take $\sigma=0.1$ for the magnetized IS. The ordered magnetic field corresponding to the magnetization in region 4 is
\begin{equation}
B'_{o,4}=\sqrt{\frac{2m_r\sigma_rc^2}{(1-\cos\theta_0)R^2\gamma_rl_r}}
\end{equation}
Using the shock jump condition, the ordered magnetic field in region 3 can be expressed as $B'_{o,3}=u_{4s}B'_{o,4}/u_{3s}$. The random magnetic field generated by the shock is $B'_{r,3}=\sqrt{8\pi\epsilon_{B,3}e'_3}$ and a part $\epsilon_{B,3}$ of the internal energy goes to the random magnetic field. Finally, the total magnetic field is obtained as $B'_{t,3}=\sqrt{B^{'2}_{o,3}+B^{'2}_{r,3}}$. With the evolution of the magnetic field, the cooling and maximum Lorentz factors are
\begin{equation}
\gamma_{c,3}=\frac{3\pi m_ec(1+z)}{\sigma_T\gamma_fB^{'2}_{t,3}t_{obs}},\ \ \ \ \ \gamma_{max,3}=\sqrt{\frac{6\pi e}{\sigma_TB'_{t,3}}}
\end{equation}
Here, $e$ is the charge of an electron, and the Thomson scattering cross section is $\sigma_T$. In the comoving frame of the region 3, the cooling Lorentz factor is $\gamma_c=6\pi m_ec/\sigma_TB^{'2}_{t,3}t'$, and $t'$ is the emitting time in the comoving frame. Considering the EATS given by Eq. (1) in this paper, the observational time $t_{obs}$ is related to the comoving time by $t_{obs}=(1+z)t'/2\gamma_f$. When $t-t_{col}\leq t_{rs}$, $B'_{r,3}\propto R^{-1}$. When $t-t_{col}>t_{rs}$, if the energy loss due to adiabatic expansion is unimportant, $B'_{r,3}\propto R^{-1.5}$, otherwise the random magnetic field will decay faster than $R^{-1.5}$.

The decay index of a total magnetic field with radius is $b=1$ before the shock crossing. The shell has adiabatic expansion after shock crossing. Its comoving width will increase as $R/\gamma_f$, where $\gamma_f$ is the bulk Lorentz factor of the emitting region. For the large-scale toroidal magnetic field, which is confined to the shock plane and axisymmetric about the jet axis, from the conservation of the magnetic flux $\Phi=B'R^2\theta_0/\gamma_f$, $b$ equals to 2. For the aligned magnetic field discussed in this paper, which is also confined to the shock plane and is latitude circles on the jet, and from conservation of the magnetic flux $\Phi=2B'R^2\theta_0/\gamma_f$, the decay index is also 2 after the shock crossing.

For simplicity, we take the adiabatic index of region j as $\hat{\Gamma}_{j}=4/3\ (j=2,3)$ before the shock crosses the shell. After the shock has crossed the shell, the total number of electrons in region 3 is $N_r=m_r/m_p$. The volume of region 3 evolves as $V'_3\propto R^3$ and the number density of region 3 is $n'_3=N_r/V'_3$. After the shock has crossed the shell, no more internal energy is generated, and the adiabatic loss is considered. The total internal energy in region 3 is $E'_3=E'_3+dE'_{ad}$, with the adiabatic energy loss $dE'_{ad}\propto-dV'_3/V^{'\hat{\Gamma}_3}_3$. The internal energy density then is $e'_3=E'_3/V'_3$. The adiabatic index evolves as follows:
\begin{equation}
\hat{\Gamma}_3=\frac{4E'_3+5m_rc^2}{3(E'_3+m_rc^2)}
\end{equation}
The minimum Lorentz factors of electrons after shock crossing can be obtained using Eq. (3). After shock crossing, the Lorentz factors of all electrons evolve adiabatically, and then $\gamma_{c,3}\propto\gamma_{m,3}$. The treatment of the FS region is similar to the above process for the RS region, so we will not repeat it here.

\subsection{Radiation and Polarization}
As discussed in the above section, the magnetic field will decay with radius. The fast-cooling synchrotron emission in a decaying magnetic field can reproduce the Band spectrum of the GRB prompt phase (Uhm \& Zhang 2014). The energy spectrum of electrons will have an asymptotic value of $\tilde{p}=(6b-4)/(6b-1)$ in the range of $\gamma_c<\gamma_e<\gamma_m$, where $b$ is the decay index of the magnetic field with radius ($B'_{t}\propto R^{-b}$). Here, for simplicity, we take the energy spectrum of electrons in such a decaying magnetic field as follows:
\begin{equation}
N(\gamma_e)\propto\begin{cases}
\gamma_e^{-\tilde{p}}, & \text{$\gamma_c<\gamma_e<\gamma_m$}, \\ \gamma_e^{-(p+1)}, & \text{$\gamma_e>\gamma_m$},
\end{cases}
\end{equation}
where $p$ is the power-law index of injected electrons.

The magnetic field configuration (MFC) in the magnetized IS model is mixed. The ordered part is carried out from the central engine, and the random part is generated by shocks. Here we take the aligned, ordered part as an example (Spruit et al. 2001; Lan et al. 2019). The observational angle $\theta_V$ is taken to be 0 rad, which will not affect our main results with the MFCs discussed in this paper (Lan et al. 2020). The concrete calculating formula for the polarization evolution with a mixed magnetic field can be found in Lan et al. (2019). The local PD $\pi_0$ in an ordered magnetic field can be expressed as $\pi_0=(1-m)/(5/3-m)$, with $f_\nu\propto\nu^m$.

\subsection{Numerical Results}
The fixed parameters we take are as follows: $m_r=m_s=10^{22}\ g$, $\gamma_r=1000$, $\gamma_s=50$, $l_r=l_s=1\ s\times c$, $l_{sep}=10\ s\times c$, $\sigma_r=\sigma_s=0.1$, $\theta_0=0.1$ rad, $p_{2}=p_{3}=2.8$, $\epsilon_{e,2}=\epsilon_{e,3}=0.1$, and the fraction of the accelerated electrons $f_{2}=f_{3}=0.008$. Also, $p_2$ ($p_3$) is the power-law index of injected electrons in the FS (RS) region. The orientation of the aligned magnetic field is set to be $\delta_{a,2}=\delta_{a,3}=\pi/6$. The source is assumed to be at a redshift of 0.1. A flat universe with $\Omega_M=0.27$, $\Omega_\Lambda=0.73$, and $H_0=71{\rm \,km\,s^{-1}\,Mpc^{-1}}$ is assumed. In our calculation, $\theta_{max}$ is always smaller than $\theta_0$. We consider two sets of parameters. For parameter I, we take $\epsilon_{B,2}=\epsilon_{B,3}=10^{-3}$. For parameter II, we take $\epsilon_{B,2}=\epsilon_{B,3}=0.3$. A parameter $\xi_{B,j}\equiv B'_{o,j}/B'_{r,j}$ ($j=2,3$) is defined (Lan et al. 2019).

With the above parameters, we then calculate the time-evolved and energy-dependent polarization, including both PD and PA.
Fig. 1 shows the time evolution of polarization. The observational frequency is taken as $\nu_{obs}=200$ keV. PD increases slightly before the RS crossing and decays all the way after it for parameter I, while it decays all the way for parameter II. After the shock crossing, that is, roughly after the peak of the light curve, PD decays for both parameter sets. The emission stops when the shock crosses the shell because the cut frequency is lower than the observational frequency after shock crossing for both parameter sets (Kobayashi 2000). Therefore, the emission will be gradually dominated by high-latitude radiation. And with the increasing time, the emission region moves toward higher latitude. Hence, higher latitude emission will be less polarized here. The only difference in parameters I and II is their energy equal participation parameter of the random magnetic field, which is larger for parameter II. Hence, the total magnetic field of parameter II is larger, leading to a higher flux for parameter II. Since the ordered component of the magnetic field is the same for the two parameter sets and $\xi_B$ is larger for parameter I, the PD of parameter I will be larger than that of parameter II (Lan et al. 2019). PA is always a constant because the direction of the ordered component of the magnetic field is fixed.

\begin{figure}
\includegraphics[angle=0,scale=0.3]{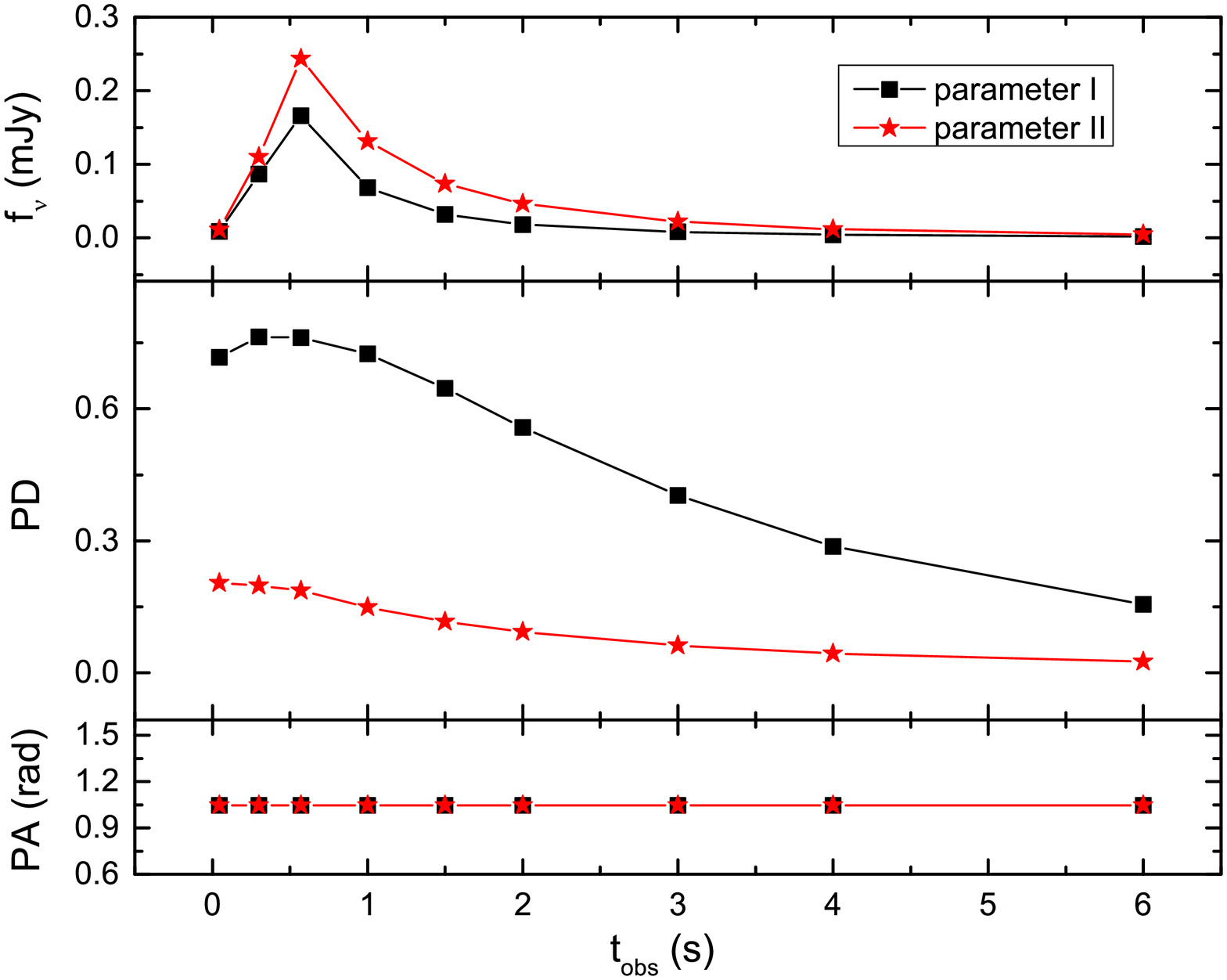}
\caption{Time-evolved polarization of the magnetized IS model. The light curves and PD and PA curves are shown in the top, medium, and bottom panels, respectively. The symbols are our calculating points, with black diamonds for parameter I and red stars for parameter II. } \label{fig1}
\end{figure}

In our calculation, MFC, the size of the $1/\gamma_f$ cone, the jet half-opening angle $\theta_0$, and the observational angle $\theta_V$ are all fixed during the time evolution. We have done some numerical experiments and define a $\tilde{f}$ parameter\footnote{Parameter $\tilde{f}$ is the flux ratio of emission within the $1/\gamma_f$ cone to the outside.} (Lan \& Dai 2020)
\begin{equation}
\tilde{f}(t_{obs})=\frac{\int_{R_c}^{R_{max}}dF_{\nu}}{\int_{R_{col}}^{R_c}dF_{\nu}},
\end{equation}
where $R_c$ is the radius at where $\theta=1/\gamma_f$, and $R_{max}$ is the maximum radius of the emission region on EATS. Taking parameter I as an example, we calculate the peak energy $E_p$ of the $\nu f_\nu$ spectrum, $\tilde{f}$, and $PD(\theta)$. $PD(\theta)$ is the PD of the emission of electrons at the $\theta$ circle (Lan \& Dai 2020). The $E_p$ values of nine calculation points (which are at 0.045, 0.3, 0.572, 1, 1.5, 2, 3, 4, and 6 s and are shown as symbols in the figures) are 344, 134, 125, 107, 89, 83, 66, 55, and 41 keV. The evolution of $E_p$ exhibits a hard-to-soft pattern. Because the emission of the jet is dominated by the RS region with parameter I, in the following we only analyze $\tilde{f}$ and $PD(\theta)$ of the RS emission. For the first six calculation points, $\tilde{f}$ is infinity. At $t_{obs}=3$ s, $\tilde{f}$ is 0.564. For the last two points, $\tilde{f}$ is 0. We have calculated the $PD(\theta)$ at 12 different EATS values (i.e., at $t_{obs}=0.572, 1, 1.7, 2.5, 3.5, 5, 7, 9, 10, 11, 13, 15$ s) with a constant $\pi_0=0.75$. With the increase in observational time, the emission region moves gradually toward the large $\theta$ region. The $PD(\theta)$ is the same for the same $\theta$ value at different EATS values and decays toward the large $\theta$ region. The $PD(\theta)$ curves of all EATS values constitute a smooth curve. It decays fast around $\theta\sim1/\gamma_f$ and approaches an approximate value of 0.75 at small $\theta$ ($<1/\gamma_f$) and zero at large $\theta$ ($>1/\gamma_f$).

With the above results of our numerical experiments, we will interpret our numerical results of Fig. 1. For the first six calculation points, their $\tilde{f}$ is infinity and all of the emission is within the $1/\gamma_f$ cone. Because $E_p$ of the first point is larger than the observational frequency (200 keV), and the spectrum index of the emission at the observational frequency is smaller, leading to a smaller $\pi_0$. For the second point, $E_p$ is smaller than the observational frequency and the spectrum index of the emission is larger at the observational frequency, resulting in a larger $\pi_0$. Therefore, PD increases slightly for the second point compared with the first point. From the third to sixth points, PD decays obviously. Although $E_p$ decreases from 125 to 83 keV, they are all smaller than the observational frequency, and the $\pi_0$ of these points are approximately the same. Note that $\tilde{f}$ is infinity, which means the emission regions are all within the $1/\gamma_f$ cone. They move gradually to the large $\theta$ region within the $1/\gamma_f$ cone with observational time, so $PD(\theta)$ is around 0.75 at the third point and decreases sharply to about 0.4 at sixth point. This is the main reason for the PD decay from third point to sixth point. For the seventh point, its $\tilde{f}$ is 0.564, and the emission within the $1/\gamma_f$ cone contributes comparably to that from the outside. Because the PD of the jet emission will decrease with a decaying $\tilde{f}$ (Lan \& Dai 2020) and $PD(\theta)$ also decreases quickly from the sixth to seventh point, the PD of the jet emission decreases from the sixth point to the seventh point. Because $PD(\theta)$ decays shallowly from the eighth point to the ninth point, the PD of the jet emission also decreases slowly.

Intensity and polarization spectra are exhibited in Fig. 2, which are calculated at the time of the light-curve peak. For both sets of parameters, energy spectra show a Band-function-like form above the X-ray band, due to synchrotron emission in a decaying magnetic field. The PAs of both sets of parameters always remain as constant with energy. For parameter I, the ratio of the strength of the ordered magnetic field to the random magnetic field is $\xi_{B,3}=16$ for the RS region. Actually, when $\xi_B$ exceeds about 10, the total magnetic field will be dominated by the ordered component, and the PD of the synchrotron emission in such a mixed magnetic field will approach the value in a corresponding pure ordered magnetic field (Lan et al. 2019). For parameter II, the ratio is $\xi_{B,3}=0.92$ for the RS region. That value lies where PD evolves quickly with $\xi_B$ parameter (Lan et al. 2019). Because $\xi_{B,3}$ of parameter I is larger than that of parameter II and RS emission dominates over FS emission at the energy band shown in the figure for both parameter sets, the PD of the jet emission for parameter I will be larger than that of parameter II.

\begin{figure}
\includegraphics[angle=0,scale=0.3]{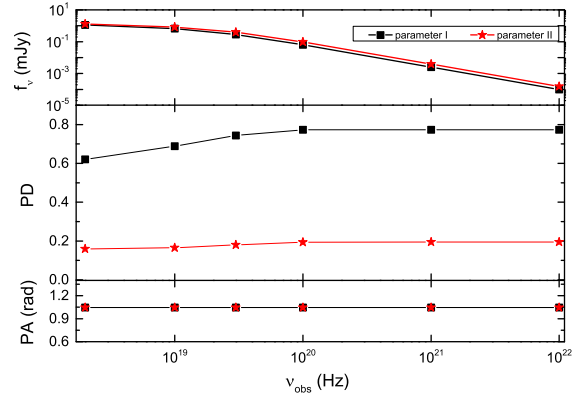}
\caption{Intensity and polarization spectra of the magnetized IS model. The energy spectrum and PD and PA spectra are shown in the top, medium, and bottom panels, respectively. The solid lines with symbols show the total results, including both the RS and FS contributions. The symbols are our calculating points, with black diamonds for parameter I and red stars for parameter II. } \label{fig1}
\end{figure}

Since most of the current polarization data in the GRB prompt phase are time-integrated, and $\xi_B$ is crucial to determining the final PD of the synchrotron emission in mixed magnetic field (Lan et al. 2019), the dependence of the time-integrated PD on $\xi_B$ is calculated and shown in Fig. 3. The time-integrated PD increases quickly with $\xi_B$ when $\xi_B<3$ and reaches a saturated value of $\sim0.7$ when $\xi_B>5$. When $\xi_B=0$, the IS is without magnetization ($\sigma_r=\sigma_s=0$) and its dynamics is given by Kobayashi et al. (1997). We take $\epsilon_{B,2}=\epsilon_{B,3}=0.3$ and other parameters are the same as their fixed value in this paper. In addition to parameters I and II and the $\xi_B=0$ case, we calculate the time-integrated PD of another four parameter sets. The equipartition parameters of magnetic fields for these four sets are $\epsilon_{B,2}=\epsilon_{B,3}=0.1277, 2.84\times10^{-2}, 1.02\times10^{-2}$, and $2.55\times10^{-3}$ and the other parameters take the fixed value. The corresponding $\xi_B$ values of these four parameter sets are $\sqrt{2}$, 3, 5, and 10. The emission is dominated by the FS for the unmagnetized case and by the RS for the other six magnetized cases. The $\xi_B$ values shown in Fig. 3 are taken as their values in the dominating emission region. Thus we take $\xi_B$ as $\xi_{B,3}$ for six magnetized cases and as $\xi_{B,2}$ for the unmagnetized case. The evolution trend of the $\xi_B-PD$ curve for the time-integrated polarization is very similar to the time-resolved results (see the inset figure of Fig. 2 in Lan et al. 2019), and it is independent of the observational energy band.

\begin{figure}
\includegraphics[angle=0,scale=0.3]{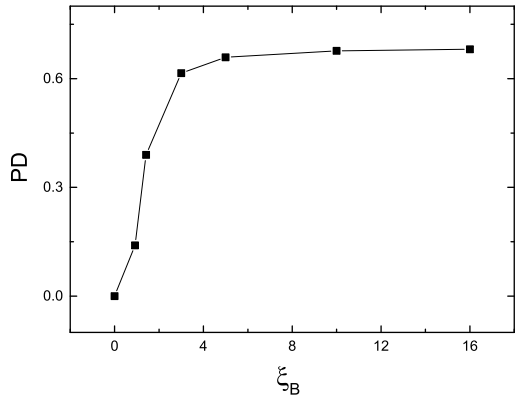}
\caption{Dependence of time-integrated PD on $\xi_B$. The black diamonds are our calculated points.} \label{fig1}
\end{figure}

\section{Conclusions and Discussion}

Up until now, several thousand GRBs have been detected. The IS model could reproduce the fast-rise and slow-decay features of the GRB prompt light curves. In this paper, the magnetized IS is considered. We reanalyze the evolution of a large-scale magnetic field and find it will decay with radius. The decay index is $b=1$ ($B'\propto R^{-b}$) before the shock crossing. Synchrotron emission in such a decaying magnetic field can reproduce the energy spectrum of the GRB prompt phase (Uhm \& Zhang 2014). In this paper, in the framework of the IS model, we discuss the polarizations of GRB prompt emission. A large PD, depending strongly on the value of $\xi_B$, will be predicted before the peak time of the light curves. Although the observational energy bands are different, the profile of the curve of the time-integrated PD on $\xi_B$ is very similar to that of the time-resolved result.\footnote{The time-resolved result is shown in Fig. 2 of Lan et al. (2019).} The unmagnetized IS with only a shock-generated random magnetic field will predict a nearly zero PD for on-axis observation (Lan et al. 2020).

For a point source, its polarization is averaged over its emitting region. We constrain our following discussion to the polarization of synchrotron emission of the on-axis observation of the top-hat jet. The electron distribution will directly affect the local PD $\pi_0$ in a local ordered magnetic field. And the steeper the spectrum of electrons, the larger the local PD will be (Rybicki \& Lightman 1979). Then in a larger region, namely a point-like region (Sari 1999), the direction of the magnetic field will be randomized. Depending on the ratio of the ordered part to that of the random part (i.e., on the value of the $\xi_B$ parameter), the PD in a point-like region will be suppressed correspondingly (Lan et al. 2019). Finally, one sums the Stokes parameters of the point-like region to get the final polarization of the jet emission. The more ordered the magnetic field in the emitting region, the larger the final PD will be. Additionally, the parameter $\tilde{f}$ will also affect the final jet PD significantly. The larger the $\tilde{f}$, the larger the final PD would be (Lan \& Dai 2020).

For the magnetized IS model, the magnetization parameter $\sigma$ might be moderately low. The ordered part of the magnetic field will hold in the jet so that MFC would not change during the burst (Deng et al. 2017). Before the RS crossing time, there are no emitting hollows on the EATS around $\theta=0$. Generally speaking, $\tilde{f}$ will not increase. Also, the $\theta$ of EATS ranges from zero to $\theta_{max}$, and $PD(\theta)$ will decrease slightly with observational time. If $\tilde{f}$ stays constant and $PD(\theta)$ decays shallowly, the PD of the jet emission will be mainly affected by the $E_p$ evolution before the shock crossing. For the hard-to-soft $E_p$ pattern, $E_p$ decreases before the RS crossing, leading to an increase in the local PD $\pi_0$. Then there might be a small bump in the PD curve before the RS crossing. For the intensity-tracking $E_p$ pattern, $E_p$ increases before the RS crossing, resulting in a decrease of local PD $\pi_0$, so the PD of jet emission will decrease all the way before the RS crossing. The evolution of the PD curve of jet emission before the RS crossing might distinguish the two $E_p$ evolution patterns.

The magnetized IS was discussed by Fan et al. (2004). In their paper, PD evolution with magnetization $\sigma$ was considered roughly. The emission is simply assumed to be proportional to the magnetic energy density. The maximum PD in an ordered magnetic field is assumed to be 0.6, corresponding to a spectral index of $m=0$ below the peak energy $E_p$. In their paper, PD increases with magnetization parameter $\sigma$. With an increase in $\sigma$, the proportion of the ordered magnetic field will increase for a fixed strength of random magnetic field. Thus $\xi_B$ will increase and then PD will increase (Lan et al. 2019).

Recently, polarization of the GRB prompt phase in a decaying magnetic field was considered by Cheng et al. (2020). They do not involve any concrete model of the GRB prompt phase (i.e., the IS or magnetic-reconnection or photosphere model). They simply consider an emitting shell with a constant velocity, which starts to radiate at $R_s$ and turns off at $R_{off}$, to generate the light curve of GRB prompt emission. The MFC is assumed to be a large-scale ordered toroidal magnetic field, and the change in PA with this MFC will only be abrupt $90^\circ$ change (Lan et al. 2016). The common features of their PD evolution are the PD plateau and steep decay part after the turn-off time of a GRB pulse $t_{off}$. The evolution of their PD curve is mainly determined by the local PD $\pi_0$, the MFC in the $1/\Gamma$ cone, and the $\tilde{f}$ parameter defined in Lan \& Dai (2020).

The PD of the magnetized IS model could increase toward the high-energy band. The concrete PD value at the high-energy band depends strongly on the parameter $\xi_B$, while for the dissipative photosphere model, the PD above the megaelectronvolt band decays to zero because of multiple scatterings (Lundman et al. 2018). Therefore, the PD spectrum of jet emission could be used to test these two models.

Polarization in the magnetic-reconnection model was discussed by Lan \& Dai (2020). Both the MFC and observational geometry are taken as the same for the magnetized IS model discussed in this paper and the magnetic-reconnection model in Lan \& Dai (2020). The ordered part of the magnetic field of the magnetic-reconnection model should decay faster than that of the magnetized IS model, because besides the expansion of the jet, its magnetic field also reconnects to release magnetic energy. Therefore, the $\xi_B$ parameter would decay faster with time for the magnetic-reconnection model. However, the ranges of the emission regions and the evolution of jet bulk Lorentz factors are also very different for these two models, and thus the evolution of $\tilde{f}$ parameters of the two models cannot be definitely determined. Therefore, it is hard to distinguish the magnetic-reconnection model and the magnetized IS model through PD evolutions of the jet.

\acknowledgments
We thank an anonymous referee for his/her helpful suggestions. We also thank Jin-Jun Geng and Xiao-Hong Zhao for useful discussions. This work is supported by the National Natural Science Foundation of China (grant Nos. 11673068, 11725314, 11903014, and 12047569). X.F.W. is also partially supported by the Key Research Program of Frontier Sciences (QYZDB-SSW-SYS005), the Strategic Priority Research Program ``Multi-wave band gravitational wave Universe'' (grant No. XDB23040000) of the Chinese Academy of Sciences, and the ``333 Project" of Jiangsu province. S.X. acknowledges the support provided by NASA through the NASA Hubble Fellowship grant No. HST-HF2-51473.001-A  awarded by the Space Telescope Science Institute, which is operated by the Association of Universities for Research in Astronomy, Inc., for NASA, under contract NAS5-26555.

\end{document}